\documentclass[12pt]{iopart}
\usepackage{epsfig}
\usepackage{iopams}

\begin{document}

\title{Extended Skyrme interaction (II): \\
ground state of nuclei and of nuclear matter}

\author{J. Margueron$^1$, S. Goriely$^2$, M. Grasso$^1$, G. Col\`o$^3$ and H. Sagawa$^4$}
\address{$^1$Institut de Physique Nucl\'eaire,
 Universit\'e Paris-Sud, IN2P3-CNRS, F-91406 Orsay Cedex, France}
\address{$^2$Institut d'Astronomie et d'Astrophysique, CP-226, 
Universit\'e Libre de Bruxelles, 1050 Brussels, Belgium}
\address{$^3$Dipartimento di Fisica, Universit\`a degli Studi and 
INFN Sez. di Milano, Via Celoria 16, 20133 Milano, Italy}
\address{$^4$Center for Mathematics and Physics, 
University of Aizu, Aizu-Wakamatsu, 965-8580 Fukushima, Japan}


\begin{abstract}
We study the effect of time-odd components of the Skyrme energy density 
functionals on the ground state of finite nuclei and in nuclear matter. 
The spin-density dependent terms,  which have been recently 
proposed as an extension of the standard Skyrme interaction,  
are shown to change  the total binding energy of odd-nuclei by 
only few tenths of keV, while the time-odd components of standard
Skyrme interactions give an effect that is larger by one order of magnitude.
The HFB-17 mass formula based on a Skyrme parametrization is adjusted 
including the new spin-density dependent terms. 
A comprehensive study of binding energies in the whole mass table of 2149 
nuclei gives a root mean square (rms) deviation of 0.575 MeV between 
experimental data and the calculated 
results, which is slightly better than the original HFB-17 mass formula.
From the analysis of the spin instabilities of nuclear matter, restrictions 
on the parameters governing the spin-density dependent terms are evaluated. 
We conclude that with the extended Skyrme interaction, the Landau parameters
$G_0$ and $G_0^\prime$ could be tuned with a large flexibility without 
changing  the ground-state properties in nuclei and in nuclear matter.
\end{abstract}




\submitto{\JPG}


\section{Introduction}

Despite many theoretical and experimental investigations, the spin and the
spin-isospin channels in either the ground and the excited states of 
nuclei are  still widely open for future study~\cite{bor84,ost92,suz99,eng99,mar02,frac07}.
It is indeed difficult to probe the spin and the spin-isospin channels
of nuclear interaction since the ground states of nuclei are non-spin 
polarized in the case of even-even nuclei and at most polarized by the 
last unpaired nucleons in odd nuclei.

The analysis of spin and spin-isospin collective modes such as  
magnetic dipole (M1) and Gamow-Teller (GT) states gives access to the
nuclear interaction in these channels. 
The Landau parameter $G_0^\prime$ has been deduced from the  analysis of the GT 
mode: a model based on Woods-Saxon single-particle states plus one-pion and rho meson
exchange interactions gives $G_0^\prime=1.3\pm0.2$ (see Ref.~\cite{ost92,suz99} 
and references therein). 
A slightly different value  $G_0^{\prime} = 1.0\pm0.1$ was  derived from observed  
GT and M1 strength distributions using the phenomenological 
energy density functionals DF3~\cite{bor84,bor06}. 
Anyway, in both cases empirical single-particle energies 
are used (that is, for states close to the Fermi energy the
effective mass $m^*/m$ is $\approx$ 1). 

Self-consistent Hartree-Fock (HF) plus Random Phase Approximation
(RPA) calculations constitute a somewhat different framework,
in which the density of states around the Fermi energy is lower 
(or, equivalently, the effective mass is about 30\% smaller). 
As compared with the empirical case, the unperturbed particle-hole
transitions have larger energies and as a consequence one needs
a smaller residual repulsive effect to fit the observed GT
peak. It is not surprising, therefore, that self-consistent 
calculations of the GT resonance, performed 
using different 
Skyrme interactions in Ref.~\cite{frac07}, point to 
$G_0^\prime \sim 0.6$. 
Skyrme interactions are characterized by
a spin-isospin $G_1^\prime$ parameter as well, and 
specific terms of the effective mean field, 
like the spin-orbit potential, may also break the simple correlation between
the GT properties and the parameter $G_0^\prime$. 
However, as widely used interaction like SLy5 have unrealistic (negative) 
values of $G_0^\prime$, it is undeniable that adding 
more flexibility to the spin-isospin part of Skyrme forces
is useful.

The spin and spin-isospin component of the nuclear interaction is also reflected
into the time-odd component of the mean field.
The properties of even nuclei give constraints to the time-even component
of the mean field while very little is known about properties of the time-odd 
mean fields.
Time-odd components of the mean field compete with pairing correlations in 
determining the odd-even mass staggering~\cite{dug01}.
Fast rotation induces time-odd components in the mean field~\cite{doba95} which
could be probed from the measurement of the dynamical moments of superdeformed 
bands.

In Ref.~\cite{marg08}, new spin-density dependent terms have been introduced
on top of standard Skyrme forces in order to remove the ferromagnetic 
instability associated with all Skyrme parameterizations~\cite{mar02}.
The new terms retain the simplicity and the good properties of Skyrme interactions
for nuclear matter and the ground states of even-even nuclei. 
However, these new terms slightly change the properties of odd systems.
This work aims at studying quantitatively the effects of these new terms  
on the ground state properties of odd nuclei,  in particular the total binding energy and the
density distribution. 
Since these terms contribute only in odd nuclei, both odd-even and odd-odd 
nuclei are considered in the present study. 
To provide an approximate  maximal estimate of the effects 
while keeping our model simple, we perform HF calculations 
with the following approximation to treat odd nuclei. 
We use the equal filling approximation, so that the time-reversal symmetry 
is not broken, but with an additional ansatz. 
As within this scheme the spin-densities would be by definition equal to zero,  
when constructing the spin-densities with the wave function of the odd nucleon, 
we assume that
the spin-up state is completely filled while the spin-down 
state is empty between the two possible spin orientations  
(or, equivalently, the opposite). 
We call this procedure for the construction of spin-densities the 
\textsl{one-spin polarized approximation} (OSPA).
The OSPA gives an upper value of the contribution of the new terms.

The article is organized as follows: 
in Sec.~II we remind the newly proposed spin-density dependent terms as in
Ref.~\cite{marg08}. In Sec.~III, we estimate how much these 
new spin-density dependent terms affect the total binding energy of odd nuclei. 
In Sec.~IV, it will be  shown that, introducing these terms in the most predictive 
HF-Bogoliubov (HFB) mass formula~\cite{BSK17}, 
the quality of the mass fit can be recovered 
with an optimal renormalization of the Skyrme parameters.
In Sec.~V we will analyze the ground state properties of infinite
nuclear matter with respect to spin-polarization and give a range for the 
parameter of newly introduced spin-density $x_3^s$.
Finally, conclusions and outlook are given in Sec.~VI.

\section{Extended Skyrme interactions}

As described in Ref.~\cite{marg08}, the new spin-density 
dependent terms added to the conventional 
Skyrme force are of the following form:
\begin{eqnarray}
V^\mathrm{s,st}(\mathbf{r}_1,\mathbf{r}_2)=
\frac{1}{6}t_3^s(1+x_3^sP_\sigma)[\rho_s(\mathbf{R})]^{\gamma_s}\delta(\mathbf{r})
+\frac{1}{6}t_3^{st}(1+x_3^{st}P_\sigma)[\rho_{st}(\mathbf{R})]^{\gamma_{st}}\delta(\mathbf{r})
\nonumber\\
\label{eq:spin}
\end{eqnarray}
where $P_\sigma=(1+\sigma_1\cdot\sigma_2)/2$ is the spin-exchange operator,
$\mathbf{r}=\mathbf{r}_1-\mathbf{r}_2$ and $\mathbf{R}=(\mathbf{r}_1+\mathbf{r}_2)/2$.
In Eq.~(\ref{eq:spin}), we have introduced the spin-density 
$\rho_s \equiv \rho_\uparrow-\rho_\downarrow$ and
the spin-isospin-density
$\rho_{st} \equiv \rho_{n\uparrow}-\rho_{n\downarrow}-\rho_{p\uparrow}+\rho_{p\downarrow}$.
Spin symmetry is satisfied if the power of the density-dependent terms $\gamma_s$ and 
$\gamma_{st}$ is even.

The total energy $E_\mathrm{TOT}$ in finite nuclei is related to the (local) energy density
$\mathcal{H}(r)$ through 
\begin{equation}
E_\mathrm{TOT}\equiv \int d^3r \; \mathcal{H}(r) \; .
\label{eq:etot}
\end{equation}
In the following, we adopt the notation of Ref.~\cite{chaba} 
where $\mathcal{H}$ is expressed as the sum of a kinetic 
term $\mathcal{K}$, a zero-range term $\mathcal{H}_0$, a density-dependent term 
$\mathcal{H}_3$, an effective-mass term $\mathcal{H}_\mathrm{eff}$, a 
finite-range term $\mathcal{H}_\mathrm{fin}$, spin-orbit and
spin-gradient terms ($\mathcal{H}_\mathrm{so}$ 
and $\mathcal{H}_\mathrm{sg}$), and eventually the 
Coulomb term $\mathcal{H}_\mathrm{Coul}$.
However, the expression for the energy density provided in~\cite{chaba} 
holds in the case of time-reversal symmetry; 
in odd nuclei, the energy density $\mathcal{H}$ acquires also a dependence 
on the spin-densities $\rho_{s}$ and $\rho_{st}$,
hereafter named $\mathcal{H}^\mathrm{odd}$, even 
without additional terms in the force depending on these
densities.
Expressions for $\mathcal{H}^\mathrm{odd}$ have been derived for instance
in Refs.~\cite{doba95,bend2002} or in Appendix~I of Ref.~\cite{flocard}.
For the reader's convenience we repeat here, in Appendix A, 
the expression for $\mathcal{H}^\mathrm{odd}$.

The additional terms~(\ref{eq:spin}) modify the standard $\mathcal{H}_3$ 
contribution to be $\mathcal{H}_3+\mathcal{H}_3^{s}+\mathcal{H}_3^{st}$,  
where the last two terms read
\begin{eqnarray}
\mathcal{H}_3^s&=&\frac{t_3^s}{12} \rho_s^{\gamma_s}
\Big[ (1+\frac{x_3^s}{2})\rho^2+\frac{x_3^s}{2}\rho_s^2 
-(x_3^s+\frac{1}{2})(\rho_n^2+\rho_p^2)-
\frac{1}{2}(\rho_{sn}^2+\rho_{sp}^2) \Big] , 
\label{eq:h3s} \\
\mathcal{H}_3^{st}&=&\frac{t_3^{st}}{12}\rho_{st}^{\gamma_{st}}
\Big[ (1+\frac{x_3^{st}}{2})\rho^2+\frac{x_3^{st}}{2}\rho_s^2 
-(x_3^{st}+\frac{1}{2})(\rho_n^2+\rho_p^2)-
\frac{1}{2}(\rho_{sn}^2+\rho_{sp}^2) \Big] ,
\label{eq:h3st}
\end{eqnarray}
with $\rho_{sn}=\rho_{n\uparrow}-\rho_{n\downarrow}$ and 
$\rho_{sp}=\rho_{p\uparrow}-\rho_{p\downarrow}$.

The additional terms to the mean field coming 
from $\mathcal{H}^{s}_3$ and $\mathcal{H}^{st}_3$ are
\begin{eqnarray}
U_q^\mathrm{s,st}&=&\frac{t_3^s}{12}\rho_s^{\gamma_s}\Big[(2+x_3^s)\rho-(1+2x_3^s)\rho_q\Big]
+\frac{t_3^{st}}{12}\rho_{st}^{\gamma_{st}}\Big[(2+x_3^{st})\rho-(1+2x_3^{st})\rho_q\Big].
\label{eq:uq}
\nonumber\\
\end{eqnarray}
where $q=n,p$. 

As already mentioned, the additional contributions to the mean field are 
zero in even-even nuclei. Since most of the Skyrme interactions are adjusted on
(few) even-even nuclei, it is thus possible to add for these interactions the 
new terms~(\ref{eq:spin}) in a perturbative manner.
The new four parameters $t_3^s$, $x_3^s$, $t_3^{st}$ and $x_3^{st}$ in Eq.~(\ref{eq:spin})
have been adjusted in Ref.~\cite{marg08} in order to reproduce the Landau parameters 
extracted from a G-matrix calculation in uniform matter, 
while $\gamma_s=\gamma_{st}=2$ is imposed by spin symmetry.

Some other interactions, in particular those produced by the 
Brussels-Montr\'eal group, are globally adjusted by fitting the 
properties (essentially the nuclear masses) of both even and 
odd nuclei.
In this case, when including the new spin-density terms, a global 
re-adjustment of the interaction might be necessary.

In the following, we first analyze the effect of the new 
spin-density dependent terms for a few selected nuclei using 
the SLy5 Skyrme interaction for which the new
spin-density dependent terms are added perturbatively.
Later, on the  basis of the latest BSk17 Skyrme parameter set, we study the 
global impact of our new terms on nuclear masses and show that the Skyrme 
parameters can be refitted to provide almost equivalent properties.

\section{Ground states of Ca nuclei}

According to the OSPA, we define the densities $\rho_s(r)$ and 
$\rho_{st}(r)$ in odd nuclei as 
\begin{eqnarray}
\rho_s(r)&=&\frac{1}{4\pi r^2}\sum_i \varphi_i^2(r) m_s(i) \; , \\
\rho_{st}(r)&=&\frac{1}{4\pi r^2}\sum_i \varphi_i^2(r) m_s(i) m_t(i),
\end{eqnarray}
where $m_s(i)$ and $m_t(i)$ are the spin and isospin z-component for each
single nucleon having the wave function $\varphi_i(r)$.
The last occupied state fully contributes to the spin-density.
It is then clear that the OSPA corresponds to maximizing the 
spin-density and its effects.

Note that if the time-reversal symmetry is not broken (in the filling 
approximation,  
for instance) both spin-up and spin-down states must be degenerate
and the densities $\rho_s$ and $\rho_{st}$ are zero.
Since in this work we aim at an approximate and maximal estimate and not 
at a precise prediction of the effects, we use the HF method with the OSPA
to treat odd nuclei.

All calculations in this Section are performed with the SLy5 parameterization 
for the conventional Skyrme force and the additional parameters  
for the spin-density terms given in Table~2 of \cite{marg08}.
We choose two systems: $^{41}$Ca and $^{49}$Ca. 
Being built on top of double magic nuclei, pairing is neglected for these
nuclei in this section.
Both nuclei are odd-even, then the proton spin-density is zero and the 
spin-isospin-density is equal to the spin-density for neutrons. 
For the nucleus $^{41}$Ca the spin-density is built with the 
neutron 1$f_{7/2}$ wave function while in $^{49}$Ca it is constructed with  
the neutron $2p_{3/2}$ wave function. 
The choice of these nuclei has been driven 
by the fact that one of the nuclei, $^{41}$Ca, has a spin-density which 
probes the external region of the nucleus, while for the other nucleus, 
$^{49}$Ca, the spin-density may probe more the central part.
We could not use the OSPA if the single-particle wave functions
changed when adding one or two nucleons. Therefore, as 
a necessary step to proceed with the evaluation of 
the impact of spin-density dependent terms within the OSPA, 
we have first checked that the wave function 
of the neutron state $1f_{7/2}$ does not change appreciably when passing 
from the even nucleus $^{40}$Ca to the next even isotope with $N+2$ 
neutrons, $^{42}$Ca. 
An analogous check has been made for the wave function of the 
proton state $1f_{7/2}$ in $^{40}$Ca and in the $Z+2$ isotone  
$^{42}$Ti.
Also, the spin-density calculated with the OSPA for $^{41}$Ca 
has  been compared 
with half the difference of the neutron densities of $^{42}$Ca and $^{40}$Ca,
where effects coming from the rearrangement of the deeper state are also included.
Very small differences are found, mainly in the central region. 
Analogous results are obtained for $^{48}$Ca, $^{49}$Ca and $^{50}$Ca.

\begin{figure}[h]
\begin{center}
\epsfig{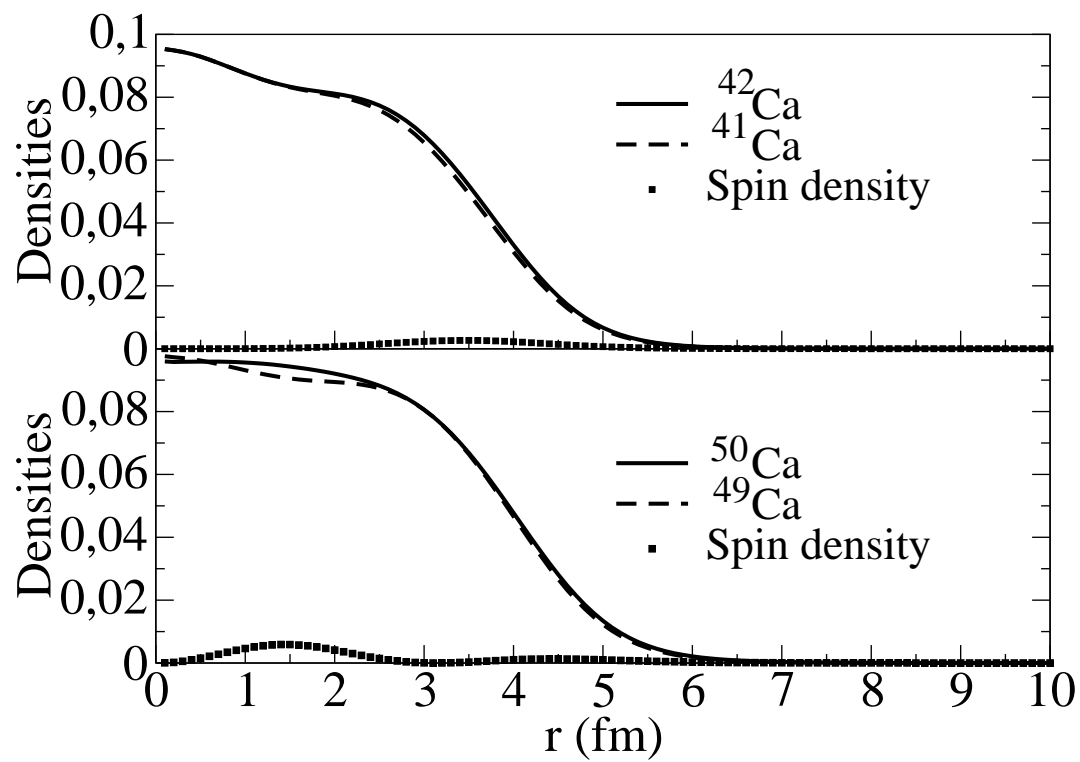}
\end{center}
\caption{Top panel: neutron density of $^{42}$Ca (solid line) and 
$^{41}$Ca (dashed line) and spin-density in $^{41}$Ca (squares). 
The densities are given in units of fm$^{-3}$. Bottom panel: the same as 
top panel, but for $^{50}$Ca and $^{49}$Ca.} 
\label{fig1}
\end{figure}

The spin-densities in $^{41}$Ca and $^{49}$Ca are plotted in the two panels of 
Fig.~\ref{fig1} together with the neutron densities of both 
nuclei. 
For comparison, the neutron densities of the nearest 
even-even nuclei $^{42}$Ca and $^{50}$Ca are also displayed. 
The spin-densities and the neutron densities 
have been calculated in the two odd-even  systems.
Two independent calculations have been actually performed with and 
without the new spin-density dependent terms~(\ref{eq:spin}) 
and only negligible differences 
have been found so that they  are not appreciable in the figures.

Since the contributions to the total energy $E_\mathrm{TOT}$
coming from the spin-density dependent 
terms~(\ref{eq:h3s}) and (\ref{eq:h3st}) are  proportional 
to the square of the spin densities represented in Fig.~\ref{fig1},
these contributions are expected to be negligible as 
compared to the usual density dependent term $\mathcal{H}_3$.
To be more quantitative, we have made several calculations for the  
ground-state energies that 
are summarized in Tables~\ref{tab:etot} and \ref{tab:particle}. 
The partial contributions to the ground-state energy 
(\ref{eq:etot}) are written 
$E_\mathrm{MF}\equiv\int d^3r \; (\mathcal{H}_\mathrm{0}+\mathcal{H}_\mathrm{3}+\mathcal{H}_\mathrm{eff}+\mathcal{H}_\mathrm{fin}+\mathcal{H}_\mathrm{sg})$,
$E_\mathrm{so}\equiv\int d^3r \; \mathcal{H}_\mathrm{so}$,
$E_\mathrm{Coul}\equiv\int d^3r \; \mathcal{H}_\mathrm{Coul}$ and
$E_\mathrm{kin}\equiv\int d^3r \; \mathcal{H}_\mathrm{kin}$, using the notations of Ref.~\cite{chaba}.
In Table~\ref{tab:etot} the total energy, the mean field, the spin-orbit, the Coulomb and the 
kinetic contributions to the total energy and the single-particle energy 
(for the neutron states $1f_{7/2}$ or $2p_{3/2}$) 
are provided for the odd nuclei $^{41}$Ca and $^{49}$Ca and for the nearest even-even nuclei 
$^{42}$Ca and $^{50}$Ca. 
The calculation for the odd nuclei are performed either with or without 
the spin-density dependent interaction~(\ref{eq:spin}).
The difference between the total energy, and its contributions, without and with the corrections
$\Delta E$ is shown to be less than 50~keV in both odd nuclei $^{41}$Ca and $^{49}$Ca.
It is interesting to notice that the kinetic energy contributes to reduce the impact of the 
spin-dependent interaction~(\ref{eq:spin}).
The spin-orbit and the Coulomb energies are very weakly affected by the spin-density terms.

In Table~\ref{tab:particle} the total and separate contributions coming from the time-odd 
terms of the Skyrme interaction (see Eqs.~(\ref{eq:a1})-(\ref{eq:a5}) in Appendix A) 
are provided, in the case of the odd nuclei $^{41}$Ca and $^{49}$Ca: these contributions 
are calculated perturbatively within the OSPA. 
These terms are classified according to the standard notations and labeled as
$E_0^\mathrm{odd}$, $E_\mathrm{eff+fin+sg}^\mathrm{odd}$ and $E_3^\mathrm{odd}$ which are 
the contributions from the central, the momentum dependent and the density 
dependent terms, respectively. 
The corrections $E_0^\mathrm{odd}$ and $E_3^\mathrm{odd}$ give a dominant and 
repulsive contribution which increases the total energy while the correction
$E_\mathrm{eff+fin+sg}^\mathrm{odd}$ is smaller and attractive.
The total correction remains quite small, that is, of the order of 0.15-0.3~MeV for both nuclei. 
Note that the sign of these corrections could change from one Skyrme interaction to another, 
but such corrections in Table~\ref{tab:particle} remain larger
than $\Delta E_\mathrm{TOT}$ in Table~\ref{tab:etot}.  
From the quantitative comparison shown in Tables~\ref{tab:etot} and \ref{tab:particle}, 
we can infer, as expected, that the new spin-density dependent 
terms~(\ref{eq:spin}) modify the ground state energies of odd nuclei much smaller than those
coming from the time-odd terms of the standard Skyrme interaction.

\section{Global adjustment on the nuclear chart}

Some Skyrme interactions have been determined by fitting the parameters to 
essentially all of the available mass data and therefore are constrained to even as well 
as odd systems. 
In this case, the new spin-density dependent terms~(\ref{eq:spin}) added to the
standard Skyrme interaction may modify the quality of the fit. To study the impact of 
the new terms on the prediction of nuclear masses, we consider now the latest 
and most accurate HFB-17 mass formula (with a rms deviation of 
0.581~MeV on the 2149 measured masses of \cite{audi03}) obtained with the 
BSk17 Skyrme force~\cite{BSK17}.

\begin{figure}[h]
\begin{center}
\epsfig{file=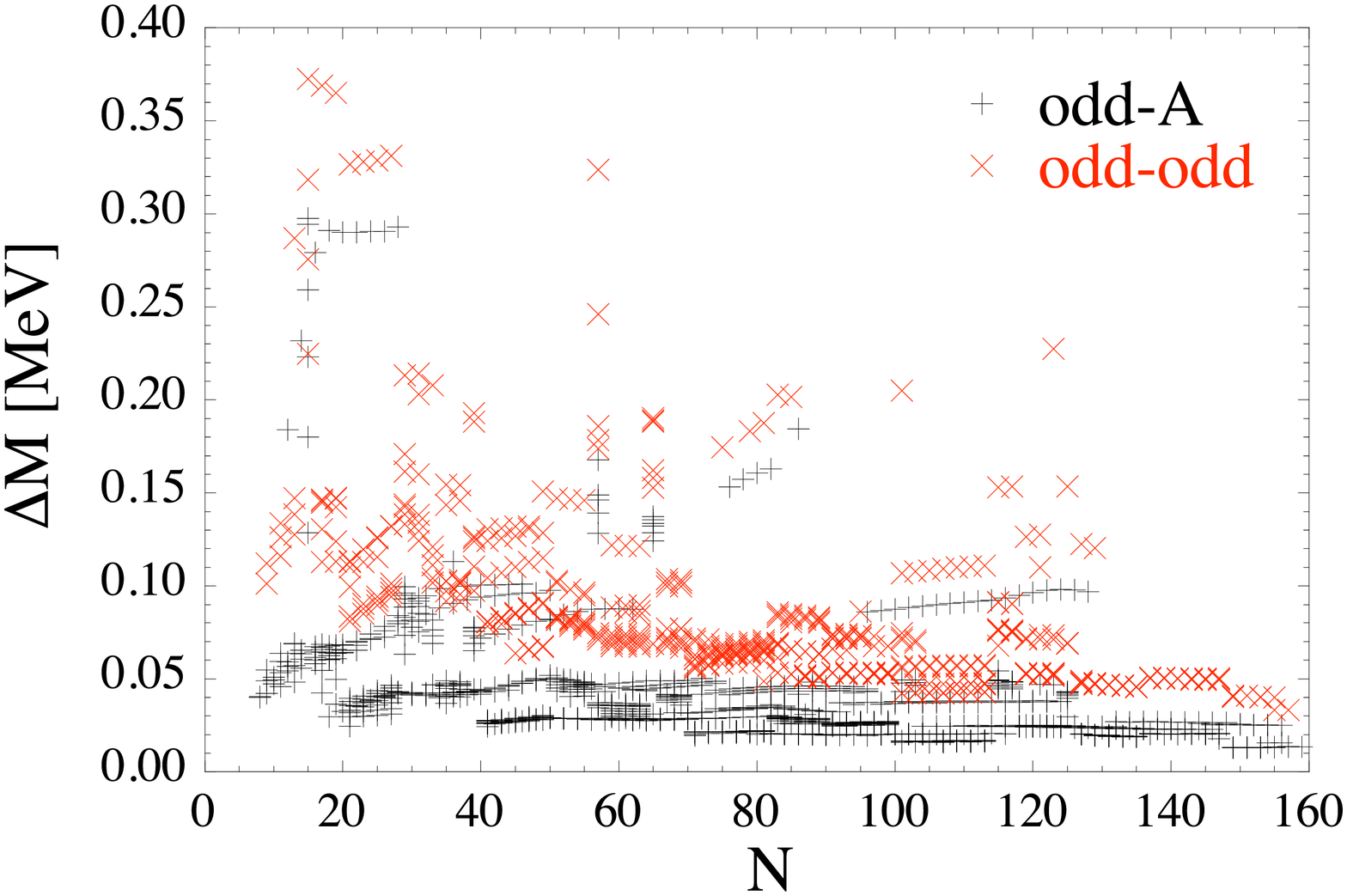,width=1.05\linewidth}
\end{center}
\caption{Difference between the nuclear mass of odd-A and 
odd-odd nuclei obtained with the BSk17 force with and without the additional 
spin-density dependent terms. The calculation is made here 
assuming spherical symmetry.
}
\label{fig2}
\end{figure}

If we consider the parameters of the additional spin-density dependent 
interaction~(\ref{eq:spin}) determined in Ref.~\cite{marg08}, 
namely $t_3^s=2\times10^4$~MeV~fm$^4$, 
$t_3^{st}=1.5\times10^4$~MeV~fm$^4$, $x_3^s=-2$, $x_3^{st}=0$ and 
$\gamma_s=\gamma_{st}=2$, we find that the impact of the new terms on 
nuclear masses are relatively small, as already discussed in Sect.~III. 
Fig.~\ref{fig2} shows the mass difference  obtained by a spherical HFB calculation 
when the spin terms are added or not. 
The nuclear masses of odd-$A$ and odd-odd nuclei are globally increased by a 
value of the order of 100~keV. 
For light nuclei this correction is the largest and can reach at most 350~keV. 
As already shown in Ref.~\cite{marg08}, the spin-density terms are repulsive and 
leads to an increase of the rms deviation with respect to all the 2149 measured 
masses from 0.581~MeV to 0.591~MeV, keeping the good quality of  the mass fit.
Deterioration can potentially be avoided if the force parameters are re-adjusted  
to the nuclear mass data.

We have  determined a new BSk17st Skyrme force which essentially 
corresponds to the BSk17 force, but for which the Skyrme as well as the pairing 
parameters have been slightly renormalized by a new fit on the whole set of 
mass data. 
The new parameter set has been built with the idea of evaluating 
a maximum effect of the new spin-density dependent term (in the
same spirit of the OSPA discussed above). The values 
adopted for the parameters are $t_3^s=4\times10^4$~MeV~fm$^4$, 
$t_3^{st}=3\times10^4$~MeV~fm$^4$, $x_3^s=-0.5$, $x_3^{st}=0$ and 
$\gamma_s=\gamma_{st}=2$. $t_3^s$ and $t_3^{st}$ are twice larger
than in Ref.~\cite{marg08} and this leads to Landau parameters
$G_0=-0.03$ and $G_0^{\prime}=0.99$ in symmetric matter at saturation 
density, whereas the parameter set of Ref.~\cite{marg08} is associated 
with lower values, namely $G_0=-0.36$ and $G_0^{\prime}=0.75$. 
As explained in the Introduction, the value of $G_0^\prime$ may
be considered large in keeping with the effective mass 0.8 of
BSk17, yet still quite acceptable for the purpose of the present
study. 
With a value of $t_3^s$ twice larger than BSk16st, the value of $x_3^s=-2$ 
in Ref.~\cite{marg08} has been modified consistently to be $-0.5$ 
to keep the contribution to the Landau parameter $G_0$ in spin-saturated 
infinite neutron matter identical to the one determined in Ref.~\cite{marg08} 
[see Eq.~(11) of Ref.~\cite{marg08}]. 
As far as the parameter $x_3^{st}$ is concerned, there is so far no constraint 
that could guide us in fixing its value. 
For this reason the zero value was assumed in Ref.~\cite{marg08}. 
However, as shown in Sec~V, the value of $x_3^{st}$ influences the stability 
of the partially polarized neutron matter, i.e., the lower its value, 
the higher the barrier between the $S=0$ and $S=1$ configuration. 
For this reason, as discussed in Sec.~V, the value of $x_3^{st}$ is set to $-3$.

The strategy of the mass fit is the same as the one described in 
Ref.~\cite{BSK17,cha08}. In particular, the mass model is given from deformed 
HFB and the pairing force is constructed from the microscopic pairing gaps 
of symmetric nuclear matter and neutron matter calculated from realistic 
two- and three-body forces, with medium-polarization effects included. 
To accelerate the fit, a first estimation of the energy gained by deformation
is performed and subtracted to the experimental masses. 
A first series of parameters are then obtained from the comparison of spherical 
HFB mass model with the corrected experimental masses.
The corrections due to deformation are then reevaluated and a new fit is performed.
This fast procedure is repeated until convergence.
The isoscalar effective mass $m^*_s/m$ is constrained to 0.80 and the  
symmetry energy at saturation $J$ to be 30~MeV in order to reproduce at best the 
energy-density curve of neutron matter~\cite{fp81} from 
realistic two- and three-nucleon forces. 
Note that the parameters of the additional spin-density interaction 
are not fitted in this procedure.

The final force parameters labeled BSk17st and resulting from a fit to 
essentially all mass data are given in Table~\ref{tab:BSK}. It can be seen 
that there is little difference between the parameters  of the BSk17 and
BSk17st forces; note that the parameters 
of the rotational and vibrational corrections~\cite{cha08},
are identical for both forces. 
The same holds for the parameters of infinite matter with the 
incompressibility coefficient $K_v=241.7$~MeV, 
the volume energy coefficient $a_v=-16.053$~MeV and the 
isovector effective mass of $m^*_v/m=0.784$, as in Ref.~\cite{BSK17}.

The rms residuals for the BSk17 and BSk17st sets are compared in Table~\ref{tab:rms}.
The inclusion of the new spin-density dependent terms~(\ref{eq:spin}) which slightly deteriorated 
the accuracy of the BSk17 force leads now  even to a small improvement of the
 predictions by about 6~keV, with the BSk17st force parameters.  
Both forces have been used to estimate the mass of the 8508 nuclei with
 $8 \le Z \le 110$ and lying between the proton and neutron drip lines.
 Differences of no more than roughly $\pm 0.5$~MeV are found on the entire set.

Finally note that the nuclear binding energies remain extremely insensitive 
to the value adopted for the parameters $x_3^s$ and $x_3^{st}$. 
In particular, setting $x_3^s=-2$ instead of $-0.5$ decreases the binding energy 
by no more than 10~keV. 
Similarly, a change of $x_3^{st}$ from zero to $-3$ impact the masses by maximum 8~keV. 
Therefore, these two parameters should be constrained rather by stability conditions 
of polarized or non-polarized infinite nuclear matter, as shown in the next Section.

\section{Ground state of infinite nuclear matter}

\begin{figure}[h]
\begin{center}
\epsfig{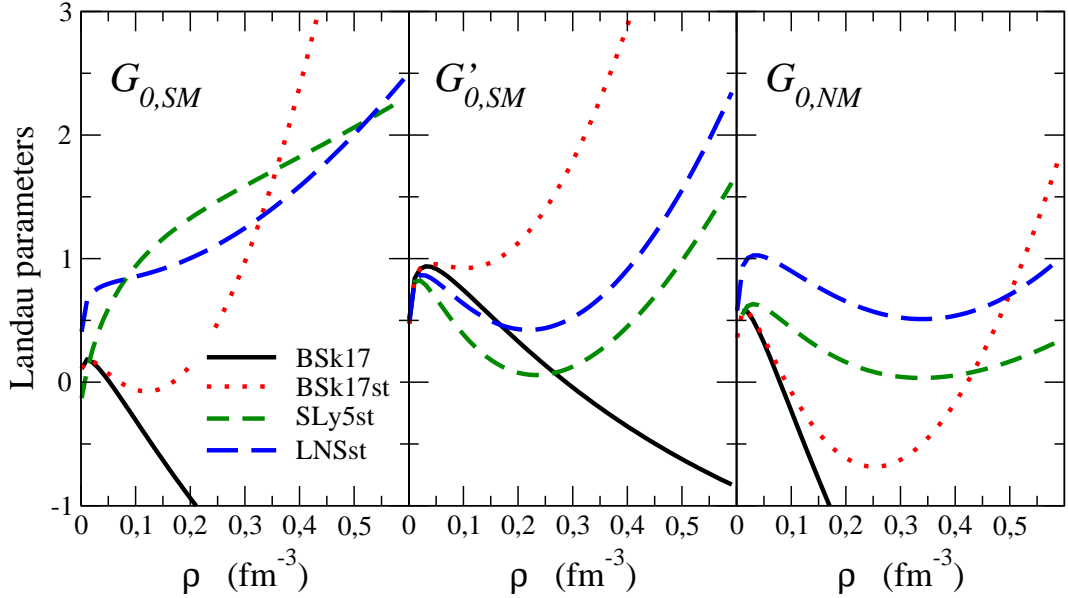}
\end{center}
\caption{Landau parameters for spin and spin-isospin channels 
as a function of density. The left panel and middle panels show $G_0$ and 
$G_0^{\prime}$, respectively, in symmetric nuclear matter and the right panel 
$G_0$ in neutron matter. The SLy5st and LNSst curves correspond to  SLy5
and LNS standard Skyrme forces  with the additional spin-density 
dependent terms, respectively~\cite{marg08}. 
The interactions BSk17 and BSk17st are described in Sect. IV.}
\label{fig3}
\end{figure}

In Ref.~\cite{marg08}, it has been shown that the new spin-density 
dependent interaction~(\ref{eq:spin})
stabilize non-polarized matter with respect to spin-fluctuations. 
As shown in Fig.~\ref{fig3}, with the new terms, the Landau parameters 
$G_0$ and $G_0^{\prime}$ remains larger than $-1$ at all densities for    
SLy5st, LNSst~\cite{LNS} (which include the new terms as parametrized 
in Ref.~\cite{marg08}) and the BSk17st forces

However, it has not been checked if the true ground state is really that of 
non-polarized matter. 
To do so, the energy for different spin-polarizations should be compared with that of 
non-polarized matter. 
This is done in the next two subsections for both symmetric nuclear matter and neutron matter

\subsection{Symmetric nuclear matter}

\begin{figure}[ht]
\begin{center}
\epsfig{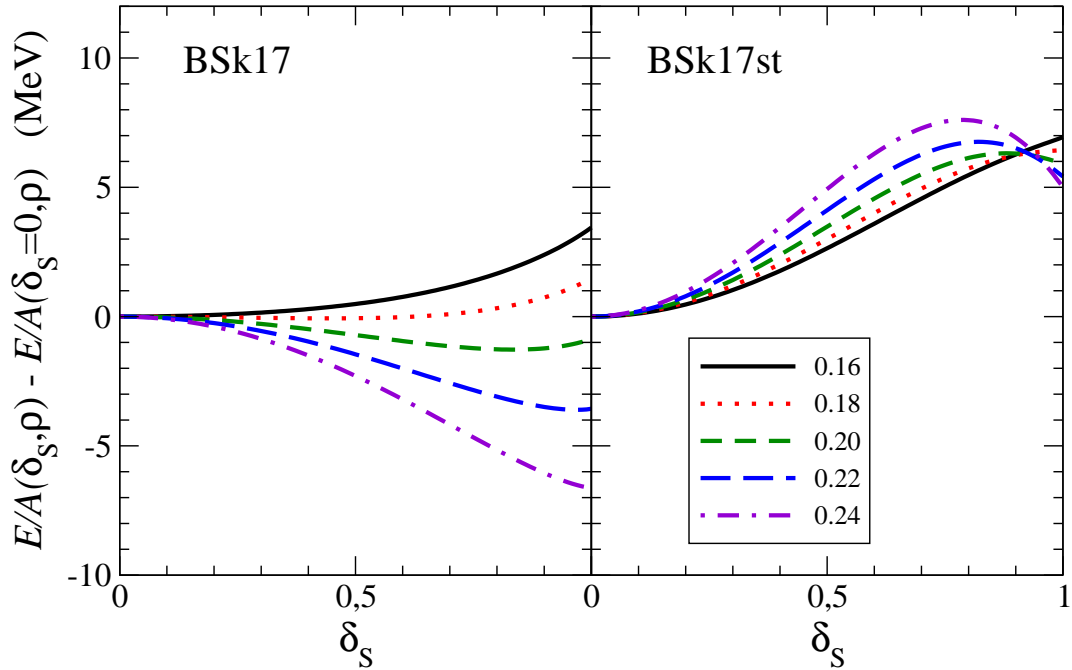}
\end{center}
\caption{Difference between the binding energies 
$E/A(\delta_S,\rho)-E/A(\delta_S=0,\rho)$ in 
symmetric matter for BSk17 and BSk17st Skyrme interaction as a function of the spin 
polarization $\delta_S=(\rho_\uparrow-\rho_\downarrow)/\rho$ 
for the different densities, as indicated in the legend (in units of fm$^{-3}$).}
\label{fig4}
\end{figure}

The difference of the binding energy of spin-polarized matter  to that 
of spin-symmetric matter,  $E/A(\delta_S,\rho)-E/A(\delta_S=0,\rho)$,  is represented in
Fig.~\ref{fig4} as a function of the polarization 
$\delta_S=(\rho_\uparrow-\rho_\downarrow)/\rho$.
In the left panel, we have represented the binding energy of the BSk17 Skyrme 
interaction without the new spin-density dependent terms~(\ref{eq:spin}).
The instability occurs between $\rho$=0.18 and 0.2~fm$^{-3}$.
At $\rho$=0.2~fm$^{-3}$ the minimum energy is obtained for a polarization
$\delta_S$=0.76.

\begin{figure}[ht]
\begin{center}
\epsfig{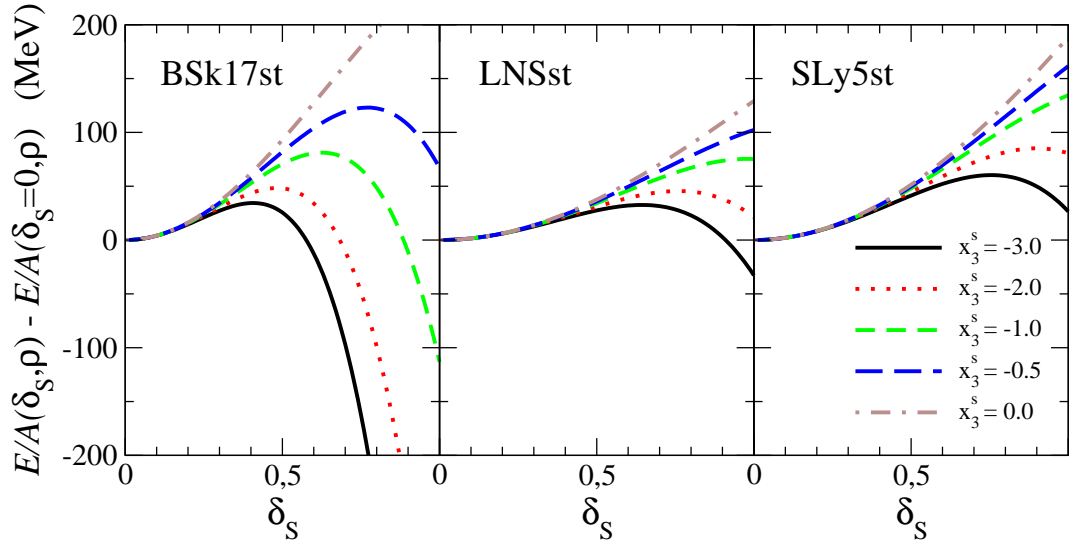}
\end{center}
\caption{Difference between the binding energies 
$E/A(\delta_S,\rho)-E/A(\delta_S=0,\rho)$ in 
symmetric matter for BSk17st, LNSst and SLy5st Skyrme interactions 
as a function of the spin polarizations and for $\rho=0.6$~fm$^{-3}$. 
A different value from 0.0 to $-3.0$ for the parameter $x_3^s$ 
is adopted for each line.}
\label{fig5}
\end{figure}

The binding energy of BSk17st which includes the spin-density dependent 
terms  is represented in the right panel of Fig.~\ref{fig4}.
As expected, the energy of non-polarized matter is convex around $\delta_S=0$,
but there is a change of convexity for large values of $\delta_S\sim 0.8$.
We have indeed observed a large influence of the parameter $x_3^s$ on the
binding energy of fully polarized matter.
In Fig.~\ref{fig5} are represented the binding energies 
$E/A(\delta_S,\rho)-E/A(\delta_S=0,\rho)$ for the three modified Skyrme interactions 
BSk17st, LNSst and SLy5st for which we changed the values of the
parameter $x_3^s$.
Its values are indicated in the legend of Fig.~\ref{fig5}.
We fixed the density $\rho$=0.6~fm$^{-3}$ to be the highest value where the nuclear Skyrme 
interaction is applied.
For values of the parameter $x_3^s$=-3 the ground state of nuclear matter is fully 
polarized ($\delta_S=1$) for the interactions BSk17st and LNSst.
Increasing the value of the parameter $x_3^s$ from -3 to 0, 
the binding energy of fully polarized matter is going up in Fig.~\ref{fig5}. 
There is then a critical value above which non-polarized matter
is the ground-state of nuclear matter.

One could obtain an estimate of this critical value by analyzing the 
contribution of the new spin-density terms~(\ref{eq:spin}) in spin-polarized 
symmetric matter. It reads
\begin{equation}
\mathcal{H}_3^s(sym.)=\frac{t_3^s}{16}\rho^2 \rho_s^{\gamma_s}
\Big[1+\frac{2x_3^s-1}{3}\delta_S^2\Big], 
\label{eq:Hs}
\end{equation}
and $\mathcal{H}_3^{st}(sym.)=0$ since $\rho_{st}=0$.
The term (\ref{eq:Hs}) is zero for the spin-symmetric matter with $\rho_s=0$ 
and is always positive for $\delta_S=1$ if one chooses $x_3^s>-1$.
It is thus clear that one necessary condition for the spin-symmetric matter to be
the absolute ground state at all densities is $x_3^s>-1$.
This is the case for the adopted value of $x_3^s=-0.5$ for BSk17st. 
Nevertheless, as shown in Fig.~\ref{fig5} for instance for SLy5st, the  
stability of spin-symmetric matter could be obtained even if $x_3^s<-1$ 
at the density $\rho$=0.6~fm$^{-3}$.   
At lower density, SLy5st is also stable, but not at higher density.

We remind that from the analysis of the Landau parameters the stability around
spin-symmetric matter requires that $x_3^s<1$ (see Eq.~(11) of Ref.~\cite{marg08}).
As a conclusion, one could adjust the parameter $x_3^s$ inside the range
$-1\lesssim x_3^s<1$.

\subsection{Neutron matter}

The case of pure neutron matter is somehow very peculiar.
The correction due to the spin-density dependent terms reads
\begin{equation}
\mathcal{H}_3^s(neut.)=\frac{t_3^s}{24}\rho^2 \rho_s^{\gamma_s}(1-x^s_3)
\Big[1-\delta_S^2\Big].
\end{equation}
It is then clear that the correction is zero for $\delta_S=0$
and also for $\delta_S=1$.
This property is related to the anti-symmetrization of the interacting nucleons.
Indeed, in fully polarized neutron matter, the quantum numbers for spin
and isospin are $S=1$ and $T=1$ while the new spin-density dependent 
interaction~(\ref{eq:spin}) act in the $L=0$ channel.
The new spin-density dependent interaction~(\ref{eq:spin}) have thus no effect 
at all in the purely spin-polarized neutron matter.
Only odd $L$ terms could play a role in  the fully polarized neutron matter.
This property has been used to provide a necessary condition to remove
the spin instabilities and lead to the condition $-5/4<x_2<-1$~\cite{kut94}.
This condition has been used in the fitting procedure of SLy5~\cite{chaba}
and it explains the robustness of the spin-symmetric ground state for this
interaction.
However if more flexibility in the Skyrme parameters is necessary, it 
might be interesting to introduce an interaction of the following form
\begin{equation}
t_5^{s}(1+x_5^s P_\sigma) \; \mathbf{k}^\prime \rho_s(\mathbf{R}) \cdot  
\delta(\mathbf{r}) \mathbf{k}\; .
\end{equation}
This $L=1$ term will not contribute to spin-symmetric matter and could be 
adjusted to fit the energy of fully polarized matter.

\begin{figure}[ht]
\begin{center}
\epsfig{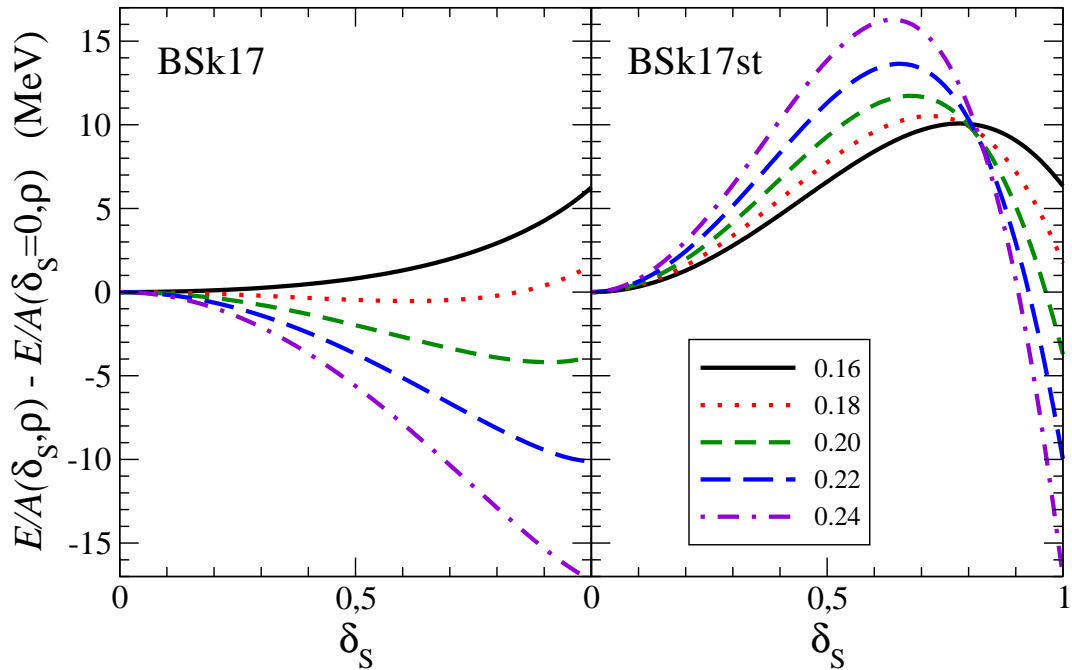}
\end{center}
\caption{Same as Fig.~\ref{fig4}, but for neutron matter.}
\label{fig6}
\end{figure}

In Fig.~\ref{fig6}, it is shown that the new spin-density terms~(\ref{eq:spin}) 
contribute to the binding energy for partially polarized matter and
tend to stabilize the state $\delta_S=0$. 
After the ferromagnetic transition, the state $\delta_S=0$ is not any more 
the absolute ground state in pure neutron matter.
However, the new spin-density  terms generate a potential barrier between the 
non-polarized and fully polarized states. 
The height of the barrier depends on the chosen parameters of the new spin
terms~(\ref{eq:spin}) as well as on the neutron matter density. 
In the right panel of Fig.~\ref{fig6}, the top of the barrier at the ferromagnetic
transition is located around the polarization $\delta_S=1/\sqrt{2}$
with a height (in MeV per nucleon) of 
\begin{equation}
\frac{1}{96}\left( t_3^s(1-x_3^s)+t_3^{st}(1-x_3^{st}) \right) \rho^3 \;.
\label{eq:bar}
\end{equation}
So, the higher the density, the higher the barrier. 
This barrier height is always positive since the curvature of the binding
energy around $\delta_S=0$ is related to the Landau parameter $G_{0,NM}$ in neutron matter
which is larger than -1 at all densities (see Fig.~\ref{fig3}).
For the BSk17st force, at $\rho\simeq \rho_f=0.19$~fm$^{-3}$ this barrier amounts 
to about 10~MeV per nucleon.

From Eq.~(\ref{eq:bar}), it can also be seen that the parameter $x_3^{st}$ influences 
the height of the barrier. 
The lower $x_3^{st}$, the larger the barrier.
For BSk17st, we set $x_3^{st}=-3$ to get a barrier above the non-polarized ground 
state of the order of 10~MeV per nucleon.
This condition is chosen with respect to the theoretical 
predictions~\cite{fan01,vid02a,vid02b,bom06} that nuclear matter is spin-symmetric 
up to reasonable high densities (see for instance discussion in the introduction 
of Ref.~\cite{marg08}).
With a barrier height of the order of 10~MeV per nucleon, newly born neutron stars 
with typical temperature going from 1 to 5~MeV might not be the site of a ferromagnetic 
phase transition.
The transition towards a non-polarized cold neutron star shall then be stable and
the remaining neutron star spin-symmetric.
Notice however that despite the theoretical predictions that dense matter is not 
spin-polarized~\cite{fan01,vid02a,vid02b,bom06},
there are no strong evidences against the occurrence of ferromagnetic phase transition 
from observation of neutron stars.
Indeed, a spin-polarized phase in the core of neutron stars might induce the very huge 
magnetic fields 10$^{15-16}$~G yet unexplained that have been proposed as the 
driving force for the braking of magnetars~\cite{hae96}.

\section{Conclusions}

The occurrence of the spin instability beyond the saturation density is a common feature 
shared by different effective mean-field approaches such as Skyrme HF, 
Gogny HF~\cite{marg01} or relativistic HF~\cite{pilar}.
The analysis of the spin component of the Skyrme interaction as well as its extensions 
might thus guide us to a wider understanding of the spin channel in general for nuclear 
interaction.
There are many reason for looking at this channel. For instance 
for its competition with pairing correlations in the odd-even-mass staggering~\cite{dug01}, 
for rotating superdeformed nuclei~\cite{doba95}, 
for a better description of GT response, 
and for all applications in astrophysics such as for instance predictions of $\beta$-decay 
half-lives of very neutron rich nuclei 
produced during the \textsl{r}-process nucleosynthesis~\cite{bor06},
reliable calculation of neutron star crust properties such as ground-states and 
collective motion~\cite{gra08},
for 0$\nu-$ and 2$\nu$ double beta decay processes, 
for URCA fast cooling, 
and also for neutrino mean free path in proto-neutron stars.

In this paper we have carefully analyzed the ground state properties of finite nuclei and
infinite matter obtained by the extended Skyrme interactions with the spin-density terms 
proposed in Ref.~\cite{marg08}.
In finite even nuclei, the new spin-density interaction~(\ref{eq:spin}) are simply zero 
and from the OSPA we have shown that these terms has only negligible contributions
to the ground-state of odd nuclei. 
These results has been obtained either by introducing the new spin-density dependent 
terms in a perturbative way to existing Skyrme interactions such as SLy5st
or performing a global adjustment of the parameter set on the nuclear chart.
A new mass formula HFB-17st adjusted in the whole isotope chart (2149 nuclei) is 
obtained with the rms deviation of about 575~keV.
From the analysis of the ground state of nuclear matter, a range for the parameter 
$x_3^s$ is restricted to $-1\lesssim x_3^s<1$ in order to stabilize the 
spin-symmetric matter.
The case of neutron matter is also discussed and it is shown that the new 
terms~(\ref{eq:spin}) with the relative angular momentum $L=0$
have no contribution to fully polarized neutron matter.
Thus, it has been shown that by using the extended Skyrme interactions~\cite{marg08}, 
the Landau parameters $G_0$ and $G_0^\prime$ could be tuned to realistic values
without altering the ground-state properties in odd nuclei as well as of nuclear matter.

\ack
S.G. acknowledges financial support from FNRS.
The work is partially supported by COMPSTAR, 
an ESF Research Networking Programme and the Japanese
Ministry of Education, Culture, Sports, Science and Technology
by Grant-in-Aid for Scientific Research under
the program numbers (C) 20540277.  

{
\appendix

\section{Time-odd components in the mean field of the Skyrme interaction}

In odd nuclei, the energy density $\mathcal{H}$ acquires a dependence 
on the spin-densities $\rho_{s}$ and $\rho_{st}$~\cite{doba95,bend2002}.
Respecting the decomposition of the Skyrme energy functional proposed in
Ref.~\cite{chaba}, the components of $\mathcal{H}^\mathrm{odd}$ are
\begin{eqnarray}
\mathcal{H}_0^\mathrm{odd} &=& \frac{1}{4}t_0\Big[(x_0-\frac{1}{2})\rho_s^2-\frac{1}{2}\rho_{st}^2\Big] \label{eq:a2} \; ,\\
\mathcal{H}_3^\mathrm{odd} &=& \frac{1}{24}t_3\rho^\gamma\Big[(x_3-\frac{1}{2})\rho_s^2-\frac{1}{2}\rho_{st}^2\Big] \label{eq:a1} \; ,\\
\mathcal{H}_\mathrm{fin}^\mathrm{odd} &=& 
\frac{1}{32}[3t_1(1-x_1)+t_2(1+x_2)]\Big(\rho_{sn}\nabla^2\rho_{sn}+\rho_{sp}\nabla^2\rho_{sp} \Big)
\nonumber\\
&&+\frac{1}{32}[t_2x_2-3t_1x_1]\Big(\rho_{sn}\nabla^2\rho_{sp}+\rho_{sp}\nabla^2\rho_{sn} \Big)
\label{eq:a3} \; ,\\
\mathcal{H}_\mathrm{eff}^\mathrm{odd} &=& 
\frac{1}{16}[-t_1(1-2x_1)+t_2(1+2x_2)]\rho_{s}\tau_s
+\frac{1}{16}[t_2-t_1]\rho_{st}\tau_{st} \label{eq:a4} \; ,\\
\mathcal{H}_\mathrm{sg}^\mathrm{odd} &=& 
\frac{1}{16}[t_1(1-2x_1)-t_2(1+2x_2)]\mathbf{j}^2_{s}
+\frac{1}{16}[t_1-t_2]\mathbf{j}^2_{st}, \label{eq:a5}
\end{eqnarray}
where $\tau_{s}$ ($\tau_{st}$) is the spin (spin-isospin) kinetic density energy defined as
$\tau_{s}=\tau_{\uparrow}-\tau_{\downarrow}$ 
($\tau_{st}=\tau_{n\uparrow}-\tau_{n\downarrow}-\tau_{p\uparrow}+\tau_{p\downarrow}$)
and $\mathbf{j}^2_{s}$ ($\mathbf{j}^2_{st}$) is the spin (spin-isospin) current.

}



\begin{table}[h]
\caption{\label{tab:etot} Total energy, mean field, spin-orbit, Coulomb and kinetic
contributions to the total energy (third column) and single-particle energy of 
the neutron state $1f_{7/2}$ for $^{41-42}$Ca and $2p_{3/2}$ for $^{49-50}$Ca 
calculated, for the nearest even nuclei and for the odd nuclei, without/with 
the spin-dependent terms~(\ref{eq:spin}) in the mean field.
$\Delta E$ is the difference of energy with and
without the  spin-dependent terms~(\ref{eq:spin}).}
\begin{indented}
\item[]\begin{tabular}{crrrrrr}
\br
Nucleus     & $E_\mathrm{TOT}$ & $E_\mathrm{MF}$  & $E_\mathrm{so}$& $E_\mathrm{Coul}$& $E_\mathrm{kin}$& s.p. energy \\
            &  (MeV)    &  (MeV)    &  (MeV)  &   (MeV)   &   (MeV)  &  (MeV) \\
\mr
$^{42}$Ca   & -362.591  & -1111.434 & -9.173 & 72.023 & 685.993 & -9.66  \\
$^{41}$Ca   &  -352.942 & -1081.395 & -5.259  & 72.116    & 661.596  & -9.64 \\
$^{41}$Ca with (\ref{eq:spin})
            &  -352.918 & -1081.359 & -5.259  & 72.115    & 661.584  & -9.64 \\
$\Delta E$  &     0.024 &     0.036 &  0.000  & -0.001    &  -0.012  & \\
\mr
$^{50}$Ca   & -429.654 & -1326.381 & -33.958 & 70.905 & 859.779 & -5.84 \\
$^{49}$Ca   &  -423.876 & -1305.865 & -33.639 & 71.105    & 844.523  & -5.70 \\
$^{49}$Ca with (\ref{eq:spin})
            &  -423.825 & -1305.754 & -33.634 & 71.102    & 844.461  & -5.70 \\
$\Delta E$  &     0.051 &     0.111 &   0.005 & -0.003    &  -0.062  & \\
\br
\end{tabular}
\end{indented}
\end{table}

\begin{table}[h]
\caption{Total and separate  contributions to the energy from the time-odd 
(spin symmetry breaking) terms of the 
SLy5 Skyrme interaction \cite{flocard,bend2002}.\label{tab:particle}}
\begin{indented}
\item[]\begin{tabular}{crrrr}
\br
Nucleus & $E_\mathrm{TOT}^\mathrm{odd}$ & $E_{0}^\mathrm{odd}$ & $E_\mathrm{eff+fin+sg}^\mathrm{odd}$ & $E_{3}^\mathrm{odd}$ \\
        &  (MeV) &  (MeV)  &  (MeV)   &   (MeV)    \\
\mr
$^{41}$Ca & 0.329 & 0.196 & -0.007  & 0.140  \\
$^{49}$Ca & 0.151 & 0.187 & -0.176  & 0.140 \\
\br
\end{tabular}
\end{indented}
\end{table}

\begin{table}[h]
\caption{Parameter sets for BSk17 and BSk17st: the first 15 lines give the Skyrme 
and additional spin parameters, the 16th the spin-orbit term, 
the 17th to 20th lines the pairing parameters and the last 4 lines the 
Wigner correction~\cite{cha08}. 
See the text and Ref.~\cite{BSK17} for more details \label{tab:BSK}} 
\begin{indented}
\item[]\begin{tabular}{ccc}
\br
& BSk17 & BSk17st \\
\mr
  $t_0$  & -1837.33 &-1837.19 \\    
  $t_1$  & 389.102 & 388.916\\      
  $t_2$  & -3.1742 & -5.3076\\      
  $t_3$  & 11523.8  & 11522.7\\     
  $t_3^{s}$  &0  & 40000 \\     
  $t_3^{st}$  &0  & 30000 \\     
  $x_0$  &  0.411377  & 0.410279\\   
  $x_1$  & -0.832102 & -0.834832\\    
  $x_2$  & 49.4875  & 29.0669\\     
  $x_3$  &  0.654962  & 0.655322\\   
  $x_3^{s}$  &  0 & -0.5\\   
  $x_3^{st}$  &  0 & -3\\   
  $\gamma$ &  0.3   & 0.3 \\
  $\gamma^{s}$ &  -   & 2 \\
 $\gamma^{st}$ &  -   & 2 \\
 $W_0$  &  145.885   & 146.048 \\  
$f_{n}^+$  &  1.000 & 1.000  \\
$f_{n}^-$  &  1.044 & 1.045 \\
$f_{p}^+$  &  1.055  & 1.059\\
$f_{p}^-$  & 1.050  & 1.059 \\
$V_W$ & -2.00  & -2.06 \\
$\lambda$                           & 320 & 410   \\
$V_W^{\prime}$  & 0.86   & 084\\
$A_0$                               & 28   & 28 \\
\br
\end{tabular}
\end{indented}
\end{table}

\begin{table}[h]
\caption{Rms ($\sigma$) deviations between experimental 
data~\cite{audi03} and HFB-17 or HFB17st predictions. 
The first line refers to all the 2149 measured masses $M$, the second to the masses 
$M_{nr}$ of the subset of 185 neutron-rich nuclei with $S_n \le $ 5.0 MeV, the 
third to the 1988 measured neutron separation energies $S_n$ and the
fourth to 1868 measured beta-decay energies $Q_\beta$. 
The fifth line shows the comparison with the 782 measured charge radii~\cite{ang04}. 
Note that units for energy and radius  are MeV and fm, 
respectively. \label{tab:rms}}
\begin{indented}
\item[]\begin{tabular}{ccc}
\br
&HFB-17&HFB-17st \\
\mr
$\sigma(2149~M)$ &0.581 &0.575   \\
$\sigma(M_{nr})$&0.729& 0.738  \\
$\sigma(S_n)$ &0.506&0.495 \\
$\sigma(Q_\beta)$ &0.583&0.585 \\
$\sigma(R_c)$ &0.0300&0.0302 \\
\br
\end{tabular}
\end{indented}
\end{table}

\end{document}